\documentclass[aps,showpacs,superscriptaddress,preprint]{revtex4}%

\usepackage{mathrsfs}
\usepackage{amsfonts}
\usepackage{amsmath}
\usepackage{amssymb}
\usepackage{graphicx}%
\usepackage{ulem}
\providecommand{\U}[1]{\protect\rule{.1in}{.1in}}

\begin{document}
\title{Phonon dispersion unfolding in the presence of heavy breaking of spatial translational symmetry}
\author{Fawei Zheng}
\affiliation{Institute of Applied Physics and Computational Mathematics, Beijing 100088, China}
\affiliation{Beijing Computational Science Research Center, Beijing 100084, China}
\author{Ping Zhang}
\thanks{Corresponding author. Email address: zhang\_ping@iapcm.ac.cn}
\affiliation{Institute of Applied Physics and Computational Mathematics, Beijing 100088, China}
\affiliation{Beijing Computational Science Research Center, Beijing 100084, China}

\begin{abstract}
The phonon dispersion unfolding method is useful for obtaining hidden Bloch symmetries and comparing theoretical results with experiment spectrums ({\it e.g.} inelastic neutron scattering, inelastic X-ray scattering, Raman ). In this paper, we propose a method to unfold phonon dispersions. The main advantage of this method is the ability to handle systems with heavy breaking of spatial translational symmetry. Its validity is tested by pure diamond, diamond with Si substitution, and diamond with C vacancies.
\end{abstract}
\maketitle

Since the discovery of effective energy bands in semiconductors\cite{boykin2007,boykin2007-2} and alloys\cite{popescu2010}, unfolding methods for treating the electronic band structures of the weakly periodic solid systems have been rapidly developed \cite{ku2010,ajoy2012,lee2013,rubel2014,deretzis2014,brommer2014,huang2014,zheng2015,farjam2015,medeiros,popescu2012}. Moreover, it has also stimulated the study of phonon dispersion unfolding method\cite{boykin2014,allen2013,huang2014,huang2015}. Like electronic energy bands, the phonon dispersions are also affected by translational symmetry (TS) breaking. Most usually, one has to artificially use a supercell to calculate the phonon dispersions of a TS-broken system. However, the theoretically calculated phonon dispersions of a supercell is much denser than that of the primitive cell. As a result, it can not be directly compared with the phonon dispersions detected in inelastic neutron (or X-ray) scattering experiments. Thus, as a further step, one needs to unfold the phonon dispersions into the Brillouin zone (BZ) of the primitive cell. Besides the experiment consideration, another benefit of the unfolded phonon dispersions is that the hidden Bloch symmetries in phonon polarization vectors are easy to find from the unfolded dispersion curves. In our previous work\cite{huang2014}, we proposed a general theoretical technique to unfold the electronic energy bands. Our technique is based on the group theory and is therefore generally applicable to all kinds of quasiparticle excitations in defective or (deformation- or magnetic order-induced) reconstructed solid systems in principle. Here in this paper, we show the application of this technique to phonon dispersions in typical defective systems, and generalize it to the phonon dispersions in a more wide range of materials.

Both electrons and phonons in crystals are elementary excitations of periodic atomic systems. The translation operators are commutative with the electronic Hamiltonian and phonon dynamical matrix. Thus, they obey the Bloch's theorem. The electronic wavefunction or the phonon polarization vector $\psi_{\vec{k}}(\vec{r})$ can be written as a product of one phase part and one periodic part. The vector $\vec{k}$ is the reciprocal variable in the first BZ. Note that one usually uses $\vec{q}$ instead of $\vec{k}$ in phonon polarization vectors. The vector $\vec{r}$ is the position of the electron or the atom. The nature of $\vec{r}$ for electronic wavefunction is different from that for phonon polarization vector, which results in differences in their unfolding methods. Normally, the electrons are described by quantum mechanics, due to their much stronger quantum effect than atoms. Thus, the electrons distribute in probability all over the real space. Whereas, the atoms are described by classical Newtonian mechanics. They localize and vibrate around their equilibrium positions. Therefore, the $\vec{r}$ is discrete for phonon polarization vectors.

In this paper, we use $\vec{r}_i$ and $\vec{R}_I$ to represent atom position vectors in primitive cell and super cell respectively. The ranges of the subscripts are $i = 1, 2, 3, ..., m$ and $I = 1, 2, 3, ..., mn$, suppose there are $n$ primitive cells in one supercell, and $m$ atoms in one primitive cell. The atom position vectors in primitive cell and super cell have relation $\vec{R}_I=\vec{r}_i+\vec{D}_d$. The vector $\vec{D}_d$ is the primitive cell lattice point vectors. The total number of irreducible primitive cell lattice point in one super cell is $n$, therefore the range of d is $d=1, 2, 3, ..., n$. The corresponding relation for electron is $\vec{R}=\vec{r}+\vec{D}_d$, where both super cell electron position $\vec{R}$ and primitive cell electron position $\vec{r}$ are continuum variables. In reciprocal space, the phonon wave vector for primitive and super cells are denoted by $\vec{q}$ and $\vec{Q}$. They obey the relation $\vec{q}=\vec{Q}+\vec{G}_b$, where  $\vec{G}_b$ is the reciprocal lattice point for super cell. Similar to $\vec{D}_d$, the total number of $\vec{G}_b$ is $n$ too, thus its range is $b=1,2,3,...,n$. The corresponding relation between primitive cell and super cell wave vectors for electrons is $\vec{k}=\vec{K}+\vec{G}_b$. All the wave vectors ($\vec{q}$, $\vec{Q}$, $\vec{k}$ and $\vec{K}$) are continuum variables.

Based on Bloch's theorem, we know that an electronic wavefunction in a supercell can be written as

\begin{eqnarray}\label{wavefunction}
    \psi_{\vec{K}}(\vec{R})=\phi_{\vec{K}}(\vec{R})e^{i\vec{K}\cdot\vec{R}}.
\end{eqnarray}
The dot between $\vec{K}$ and $\vec{R}$ refers to vector dot product.
The function $\phi_{\vec{K}}(\vec{R})$ is periodic in supercell lattice. It may be not a periodic function in primary cell lattice. However, we can write it as a superposition of primary cell periodic functions as follows,
\begin{eqnarray}\label{wavefunction}
    \phi_{\vec{K}}(\vec{R})=\sum_{b} c_{\vec{K}+\vec{G}_b}\phi'_{\vec{K}+\vec{G}_b}(\vec{R})e^{i\vec{G}_b\cdot\vec{R}}
\end{eqnarray}
where $\phi'_{\vec{K}+\vec{G}_b}(\vec{R})$ is periodic function in primitive cell lattice, therefore it obeys $\phi'_{\vec{K}+\vec{G}_b}(\vec{R})=\phi'_{\vec{K}+\vec{G}_b}(\vec{r}+\vec{D}_d)=\phi'_{\vec{K}+\vec{G}_b}(\vec{r})$. The parameter $c_{\vec{K}+\vec{G}_b}$ is a complex number. The square of its absolute value $|c_{\vec{K}+\vec{G}_b}|^2$ is the so-called unfolding weight. This unfolding weight for electrons can be calculated by using our electronic energy band unfolding technique \cite{huang2014,zheng2015}.

Let us consider the case of phonons. The phonon polarization vector in a supercell can be written as
\begin{eqnarray}\label{wavefunction}
    \psi^s_{\vec{Q}}(\vec{R}_I)&=&\phi^s_{\vec{Q}}(\vec{R}_I)e^{i\vec{Q}\cdot\vec{R}_I}  \,\,\,\
\end{eqnarray}
Here, we omit the phonon branch index. The integer $s=1,2,3$ correspond to $x$, $y$, and $z$ directions respectively. Similar to electronic wavefunction, the periodic function of phonon polarization vector $\phi^s_{\vec{Q}}(\vec{R}_I)$ can be written as follows,
\begin{eqnarray}
\phi^s_{\vec{Q}}(\vec{R}_I)&=&\sum_{b} c_{\vec{Q}+\vec{G}_b}\phi'^s_{\vec{Q}+\vec{G}_b}(\vec{R}_I)e^{i\vec{G}_b\cdot\vec{R}_I} \,\,\,\
\end{eqnarray}

where $\phi'^s_{\vec{Q}+\vec{G}_b}(\vec{R}_I)$ is the periodic wave function in primitive cell.
By defining $|\phi'_{\vec{Q}+\vec{G}_b}>=\phi'^s_{\vec{Q}+\vec{G}_b}(\vec{R}_I)e^{i\vec{G}_b\cdot\vec{R}_I}$ and $|\phi_{\vec{Q}}>=\phi^s_{\vec{Q}}(\vec{R}_I)$, this equation can be written as
\begin{eqnarray}
|\phi_{\vec{Q}}>&=&\sum_{b} c_{\vec{Q}+\vec{G}_b}\,|\phi'_{\vec{Q}+\vec{G}_b}>. \label{period}
\end{eqnarray}
The phonon unfolding weight is $|c_{\vec{Q}+\vec{G}_b}|^2$. The essence of our present phonon unfolding methods is to find this value.

\begin{figure}[ptb]
\begin{center}
\includegraphics[width=0.6\linewidth]{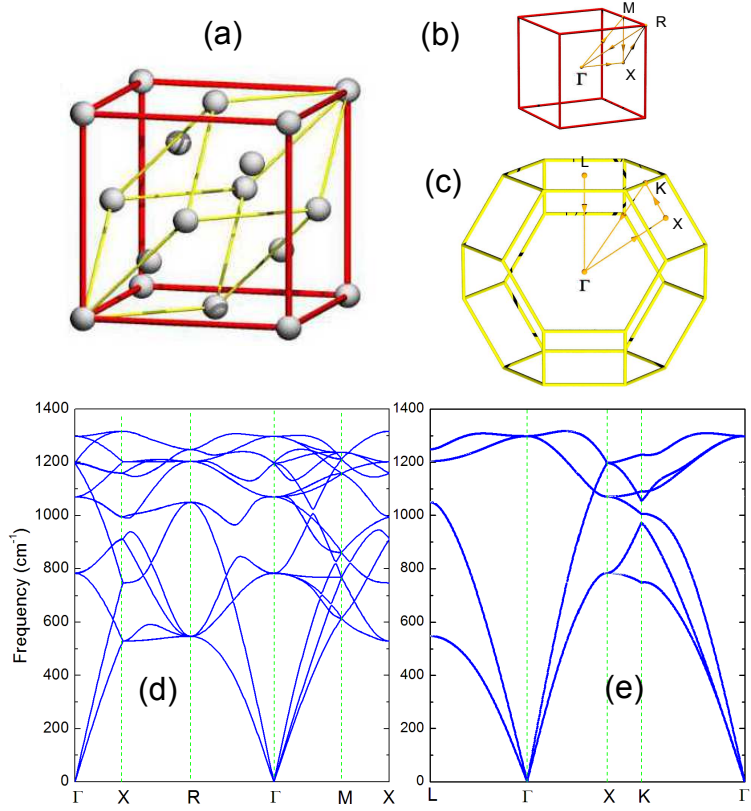}
\end{center}
\caption{(Color online)  (a) Atomic structure of diamond with its supercell (red cube cage) and primary cell (yellow parallelepiped cage). Panels (b) and (c) show the first BZs for the supercell and primitive cell, respectively, with high-symmetry points and lines. Panel (d) shows the phonon dispersions of the supercell. Panel (e) shows the phonon dispersions unfolded into the primitive-cell BZ.}
\label{D}%
\end{figure}

The supercell composes of several primary cells. If atom numbers are the same and atom positions are similar for different primary cells, then the primary cell periodic function $\phi'^s_{\vec{Q}+\vec{G}_b}(\vec{R}_I)$ is well defined, and it can be written as $\phi'^s_{\vec{Q}+\vec{G}_b}(\vec{R}_I)=\phi'^s_{\vec{Q}+\vec{G}_b}(\vec{r}_i)$. In this case, we can use the complete basis functions for primary cell to construct a projection operator. Suppose the complete basis set are $w^s_j(\vec{r}_i)\,\,\,(j=1,..,3m)$ and satisfy the following orthogonal relation:
\begin{eqnarray}
\sum_{i,s}w^{s*}_j(\vec{r}_i)w^s_{j'}(\vec{r}_i)=\delta_{j,j'},
\end{eqnarray}
The total number of basis functions is the same as the number of degrees of freedom in a primary cell, which is $3m$.  In a super cell, the primitive cell basis set can be written as  $w^s_j(\vec{R}_I)=w^s_j(\vec{r}_i+\vec{D}_d)=w^s_j(\vec{r}_i)$.
Then we define the basis set in supercell as,
\begin{eqnarray}
   |w_{j,b}>=&&\frac{1}{\sqrt{n}}w^s_j(\vec{R}_I)e^{i\vec{G}_b\cdot\vec{R}_I}\nonumber\\
   =&&\frac{1}{\sqrt{n}}w^s_j(\vec{r}_i)e^{i\vec{G}_b\cdot(\vec{r}_i+\vec{D}_d)} \,\,\,\
\end{eqnarray}
We can check that the inner product follows
\begin{eqnarray}
  <w_{j,b}|w_{j',b'}>&&=\frac{1}{n}\sum_{i,s,d}w^{s*}_j(\vec{r}_i)w^s_{j'}(\vec{r}_i)e^{i(\vec{G}_{b'}-\vec{G}_b)\cdot(\vec{r}_i+\vec{D}_d)}\nonumber\\
  &&=\delta_{j,j'}\delta_{b,b'}.
\end{eqnarray}
The projection operator is defined as
\begin{equation}
   \hat{P}_b= \sum_j|w_{j,b}><w_{j,b}|.
\end{equation}
Putting this projection operator to both sides of Eq. (\ref{period}), one has
\begin{eqnarray}
  \hat{P}_b |\phi_{\vec{Q}}> &=& \hat{P}_b\sum_{b'} c_{\vec{Q}+\vec{G}_{b'}}\,|\phi'_{\vec{Q}+\vec{G}_{b'}}> \nonumber\\
   &=& \sum_{b'} c_{\vec{Q}+\vec{G}_{b'}}\,|\phi'_{\vec{Q}+\vec{G}_{b'}}>\delta_{b,b'} \nonumber\\
   &=& c_{\vec{Q}+\vec{G}_{b}}\,|\phi'_{\vec{Q}+\vec{G}_b}>.
\end{eqnarray}
Then, the unfolding weight is
\begin{eqnarray}
  |c_{\vec{Q}+\vec{G}_{b}}|^2 &=& <\phi'_{\vec{Q}+\vec{G}_b}|c_{\vec{Q}+\vec{G}_{b}}^*c_{\vec{Q}+\vec{G}_{b}}\,|\phi'_{\vec{Q}+\vec{G}_b}> \nonumber\\
   &=& <\phi_{\vec{Q}}|\hat{P}_b \hat{P}_b|\phi_{\vec{Q}}> \nonumber\\
   &=& <\phi_{\vec{Q}}|\hat{P}_b|\phi_{\vec{Q}}>.
\end{eqnarray}
Therefore, by calculating the expectation value of projection operator $\hat{P}_b$, we can get the unfolding weight. The choice of basis functions $w^s_j(\vec{r}_i)$ is arbitrary, with the only constraint that they are unitary, complete and orthogonal to each other.
In the following calculations in this section, we choose the basis functions as
\begin{eqnarray}
w^s_j(\vec{r}_i)=\delta_{j,(s-1)m+i}
\end{eqnarray}
Then, we can calculate the inner product to be
\begin{eqnarray}
&&<w_{j,b}|\phi_{\vec{Q}}>=\sum_{I,s}\frac{1}{\sqrt{n}}w^{s*}_j(\vec{R}_I)e^{-i\vec{G}_b\cdot\vec{R}_I}\phi_{\vec{Q}}^s(\vec{R}_I)\nonumber\\
          &&\,\,\,\,\,\,=\sum_{i,d,s}\frac{1}{\sqrt{n}}w^{s*}_j(\vec{r}_i)e^{-i\vec{G}_b\cdot(\vec{r}_i+\vec{D}_d)}\phi_{\vec{Q}}^s(\vec{r}_i+\vec{D}_d)\nonumber\\
          &&\,\,\,\,\,\,=\sum_{d}\frac{1}{\sqrt{n}}e^{-i\vec{G}_b\cdot(\vec{r}_{j-[\frac{j}{m}]}+\vec{D}_d)}\phi_{\vec{Q}}^{[\frac{j}{m}]+1}(\vec{r}_{j-[\frac{j}{m}]}+\vec{D}_d)\nonumber\\
          &&\,\,\,\,\,\,=e^{i\theta(b,j)}\sum_d\frac{e^{-i\vec{G}_b\cdot\vec{D}_d}}{\sqrt{n}}\phi_{\vec{Q}}^{[\frac{j}{m}]+1}(\vec{r}_{j-[\frac{j}{m}]}+\vec{D}_d)\nonumber
\end{eqnarray}
The $e^{i\theta(b,j)}=e^{-i\vec{G}_b\cdot\vec{r}_{j-[\frac{j}{m}]}}$ is just a phase term, and it has no effect on the unfolding weight. The square bracket in $[\frac{j}{m}]$ is modulo operation. It finds the remainder after division of $j$ by $m$. From this formula, we obtain that the unfolding weight is
\begin{eqnarray}
|c_{\vec{Q}+\vec{G}_{b}}|^2 &=&<\phi_{\vec{Q}}|\hat{P}_b|\phi_{\vec{Q}}>\nonumber\\
&=&\sum_j <\phi_{\vec{Q}}|w_{j,b}><w_{j,b}|\phi_{\vec{Q}}>\nonumber\\
&=&\sum_j \left|\sum_d\frac{e^{-i\vec{G}_b\cdot\vec{D}_d}}{\sqrt{n}}\phi_{\vec{Q}}^{[\frac{j}{m}]+1}(\vec{r}_{j-[\frac{j}{m}]}+\vec{D}_d)\right|^2\nonumber\\
&=&\sum_{i,s} \left|\sum_d\frac{e^{-i\vec{G}_b\cdot\vec{D}_d}}{\sqrt{n}}\phi_{\vec{Q}}^{s}(\vec{r}_{i}+\vec{D}_d)\right|^2.
\label{method_one}
\end{eqnarray}
By using this equation, we unfolded the phonon dispersions of diamond (see below for details) as shown in Fig. \ref{D}.

The calculations were carried out within the framework of density functional theory (DFT) as implemented in
the QUANTUM ESPRESSO package\cite{qe}. The core electrons of carbon atoms are described by norm-conserving pseudopotentials\cite{nc}. The exchange correlation potential is described by the GGA of PBE-type\cite{pbe1}.
The kinetic energy cutoff for electronic wavefunction is chosen to be 70 Ry, which is well converged in our test. We use a cubic
supercell in our calculations as shown in Fig. \ref{D}(a). Each supercell (as shown by the red cage) has eight carbon atoms. The primitive
cell is also shown in Fig. \ref{D}(a) by the yellow cage. The relaxed supercell lattice parameter is 3.57 \AA, which agrees well with
the experimental results. The system is relaxed until the force on each atom is smaller than 0.01 eV/\AA. The quasi-Newton algorithm is used in the structure relaxations. In the self-consistent ground state calculations, a $13\times13\times13$
Monkhorst-Pack $k$-point setting is used in the reciprocal space integration. Then we performed the density-functional perturbation calculation with $5\times5\times5$ $q$-points to get the interatomic force constants. After that, unfolding procedure was performed to get the unfolded phonon dispersions. The resultant phonon dispersions for the supercell are shown in Fig. \ref{D}(d). And the unfolded
phonon dispersions are shown in Fig. \ref{D}(e). The atomic structure in the supercell is perfect, thus all the translational symmetries are conserved. One can see that there are fewer lines in Fig. \ref{D}(e) than in \ref{D}(d). Within our expectation, the shapes of the unfolded phonon dispersions completely agree with those calculated directly from the primitive cell.

In the presence of heavy TS breaking, the atoms in one original primitive cell may be much different with another original primitive cell. Typical examples include the doping atoms, S-W defect in graphene or carbon nanotubes, vacancies, or even liquid, etc. In such kind of systems,  the relation $\vec{R}_I=\vec{r}_i+\vec{D}_d$ is broken. The phonon polarization vector can not be written as $\phi_{\vec{Q}}^{s}(\vec{r}_{i}+\vec{G}_d)$. Also, the basis functions can not be written as $w^s_j(\vec{r}_i+\vec{G}_d)$. Instead, we can only write them as $w^s_j(\vec{R}_I)$ and $\phi_{\vec{Q}}^{s}(\vec{R}_{I})$ where the subscript $I$ runs over all the $N$ ( may not equal to $mn$ ) atoms in a super cell.  As a consequence, the whole process in method one can not be done. However, by generalizing Eq. (\ref{method_one}), now we propose another method to unfold the phonon dispersions of all these defective systems.

We write the unfolding weight [Eq. (\ref{method_one})] as
\begin{eqnarray}
|c_{\vec{Q}+\vec{G}_b}|^2 &=&\sum_{i,s} \left| A_{i,s}^{\vec{Q}+\vec{G}_b} \right|^2
\label{method_one_2}
\end{eqnarray}
with
\begin{eqnarray}
A_{i,s}^{\vec{Q}+\vec{G}_b}&=&\sum_d\frac{e^{-i\vec{G}_b\cdot\vec{D}_d}}{\sqrt{n}}\phi_{\vec{Q}}^{s}(\vec{r}_{i}+\vec{D}_d).
\label{method_one_2-1}
\end{eqnarray}
Obviously, $A_{i,s}^{\vec{Q}+\vec{G}_b}$ are the plane wave projection coefficients in supercell for each primitive cell freedom $\{i,s\}$. In a system with TS heavily broken, the $A_{i,s}^{\vec{Q}+\vec{G}_b}$ may be not possible to obtain. Here we propose that, as an approximation, the unfolding weight can also be calculated from the projection of all the degrees of freedom in a supercell to a group of plane waves. The projector operator is defined as
\begin{equation}
   \hat{P}_b= \sum_{j,s}|w_{j,s,b}><w_{j,s,b}|\nonumber
\end{equation}
with
\begin{equation}
   |w_{j,s,b}>=\frac{e^{i(\vec{G}_b+\vec{g}_j)\cdot\vec{R}_I}}{\sqrt{N}}\delta_{s,s'},\nonumber
\end{equation}
where the $\vec{G}_j$ denotes the reciprocal lattice point of primitive cell. The number of $\vec{G}_j$ used in unfolding process is finite, since the vibrational system only has finite degrees of freedom. In our calculations, we symmetrically chose finite number of $\vec{G}_j$ around (0,0,0) point. The total number of  $\vec{G}_j$ is in the same order of atom number in a primitive cell.
Then, the unfolding weight can be written as
\begin{eqnarray}
|c_{\vec{Q}+\vec{G}_b}|^2 &=&<\phi_{\vec{Q}}|\hat{P}_b|\phi_{\vec{Q}}>\nonumber\\
&=&\sum_{j,s} |B_{j,s}^{\vec{Q}+\vec{G}_b}|^2
\label{method_two}
\end{eqnarray}
with
\begin{eqnarray}
B_{j,s}^{\vec{Q}+\vec{G}_b}&=&<w_{j,s,b}|\phi_{\vec{Q}}>\nonumber\\
 &=& \sum_I\frac{e^{-i(\vec{G}_b+\vec{g}_j)\cdot\vec{R}_I}}{\sqrt{N}}\phi_{\vec{Q}}^{s}(\vec{R}_{I}).
\label{method_two-2}
\end{eqnarray}

\begin{figure}[ptb]
\begin{center}
\includegraphics[width=0.6\linewidth]{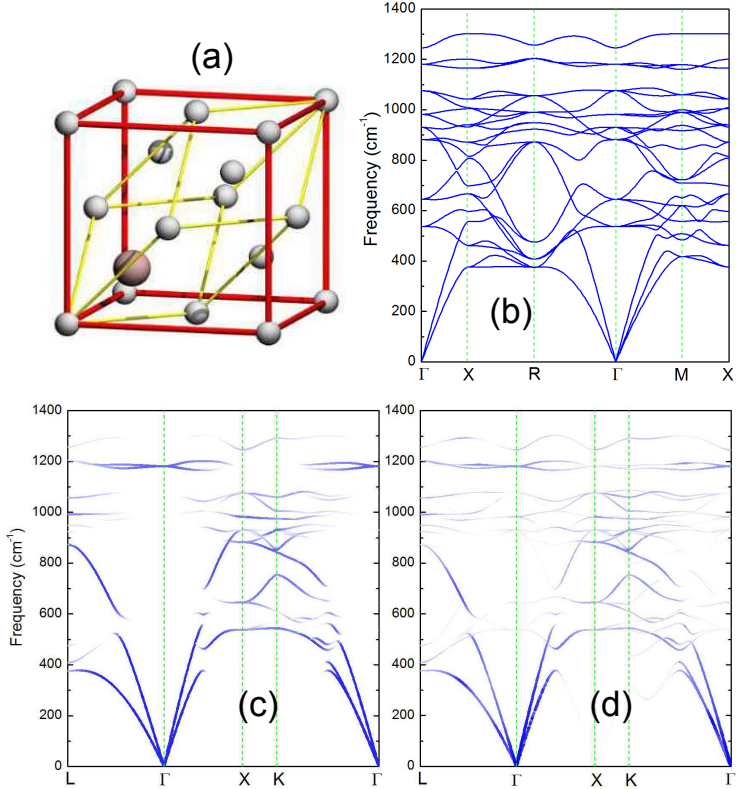}
\end{center}
\caption{(Color online) (a) Atomic structure of diamond with one C atom substituted by a Si atom. Panel (b) shows the phonon dispersions of the supercell. Panels (c) and (d) show the unfolded phonon dispersions by using the first and second methods, respectively.}
\label{D-Si}%
\end{figure}

\begin{figure}[ptb]
\begin{center}
\includegraphics[width=0.6\linewidth]{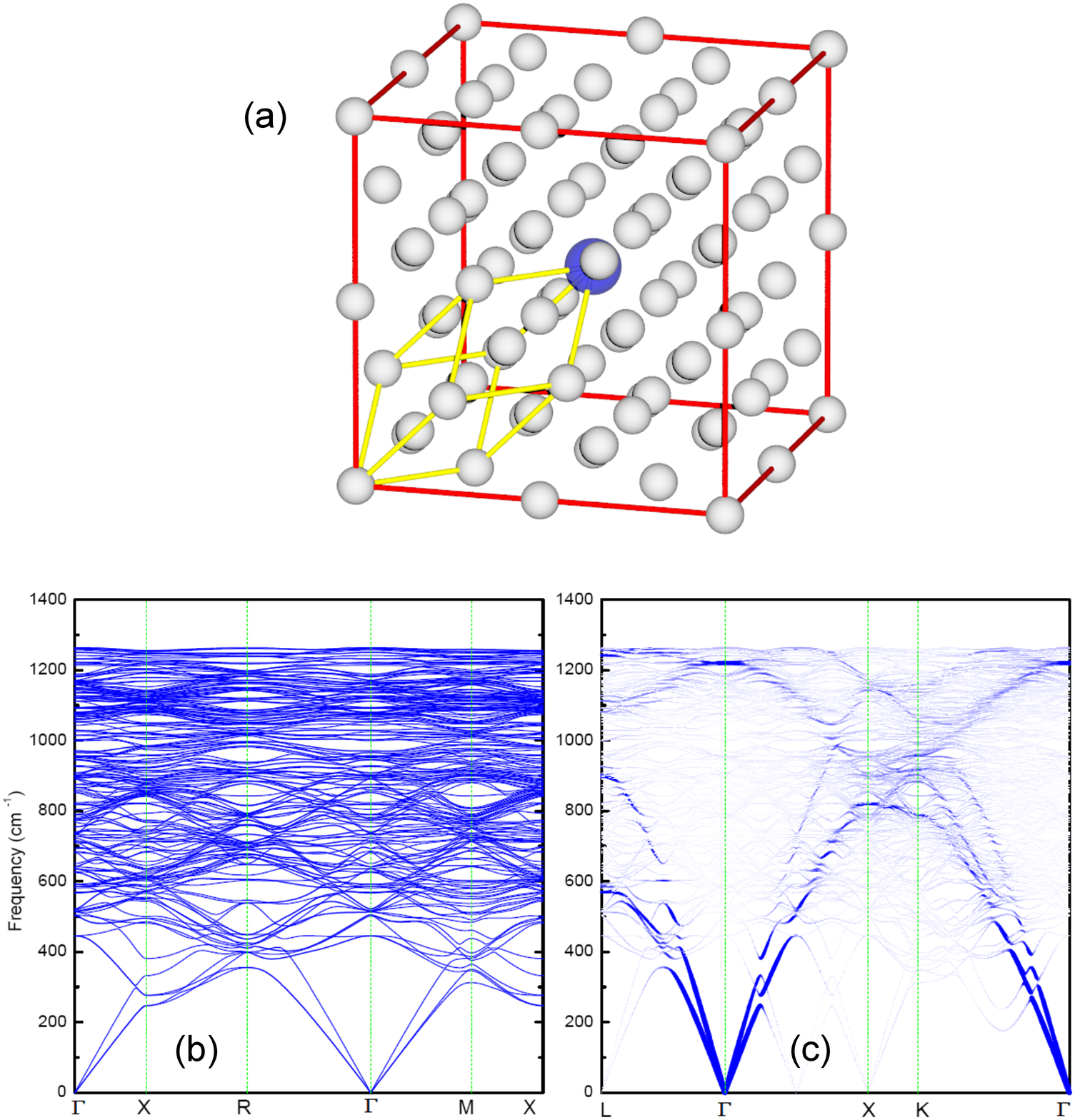}
\end{center}
\caption{(Color online) (a) Atomic structure of diamond with one carbon vacancy denoted by a blue ball. (b) Phonon dispersions of the diamond supercell with one C atom vacancy. (c) The phonon dispersions unfolded into primitive-cell BZ by using the second method.}
\label{D-Vac}%
\end{figure}

To check the validity of this general method, we performed the phonon unfolding procedure on diamond with Si substitution. This example shows the effect of TS breaking. The system is relaxed until the force on each atom is smaller than 0.01 eV/\AA.  In the self-consistent ground state calculations, a $8\times8\times8$ Monkhorst-Pack $k$-point setting is used in the reciprocal space integration. Then we performed the density-functional perturbation theory with $3\times3\times3$ $q$-points to get the interatomic force constants. The other parameters used here are the same as those used in the perfect diamond calculations. The phonon dispersions in the supercell are shown in Fig. \ref{D-Si}(b). They are more complex and disordered than Fig. \ref{D}(d). The unfolded phonon dispersions are shown in Figs. \ref{D-Si}(c,d), calculated by using method one and method two, respectively. We used  a $7\times7\times7$ plane-wave basis set in the unfolding method two.  Comparing with Fig. \ref{D}(e), the unfolded phonon dispersions of the doped system have broken points and darkness in a variety, which originate from the TS breaking. Figures \ref{D-Si}(c) and \ref{D-Si}(d) are quite similar. It means that the unfolded phonon dispersions from method two recovers the main features of that from method one. Thus method two is valid.

In order to further explore more variety in usefulness of method two, we calculated the unfolded phonon dispersions of  diamond with a carbon vacancy. In this case, method one cannot be applied because the atomic position correspondence between different primitive cells are absent. The atomic structure is shown in Fig. \ref{D-Vac}(a). We used a large supercell with 63 carbon atoms and one carbon vacancy. The carbon vacancy are shown by the large blue ball. The supercell and primitive cell are shown by the large red and small yellow cages, respectively. The calculated supercell phonon dispersion curves are shown in Fig. \ref{D-Vac}(b). They are very dense and complex. The corresponding unfolded phonon dispersions are shown in Fig. \ref{D-Vac}(c). We see that the phonon spectrum is fuzzy lines. The shading of these lines show a pattern which agree with the perfect diamond phonon dispersions in Fig. \ref{D}(e).

In summary, we proposed two methods to unfold phonon dispersions. Our method one is equivalent to the method proposed by Boykin {\it et al.} [Phys. Rev. B \textbf{90} 205214 (2014)]. This method is applicable to systems whose atom positions are not much changed. Typical examples include alloys, surface reconstruction of semiconductors, and magnetic orders. In these systems, the atomic position correspondence between different primitive cells are available, thus the phonon dispersion can be unfolded by using the basis functions in primitive cell rigorously.  When the TS is more heavily broken, we can not find the atomic position correspondence between different primitive cells. As a result, method one can no longer be employed. Then, we proposed more general method two to unfold phonon dispersions in a more wide range of systems. Within this method we project the phonon polarization vector to a set of plane waves, and obtain the unfolding weight from the projection weight. The methods were checked by three examples. The calculation results of perfect diamond and diamond with Si substitution show the validity of method two. It reproduces the same results of method one. The calculation results of diamond with a carbon vacancy shows that the second unfolding method is applicable to the systems where the first method is no longer valid. The phonon unfolding methods may increase our ability on analysing inelastic neutron diffraction experiments, and help us to gain deeper knowledge of vibrational properties of translational symmetry broken systems.

This work was supported by Natural Science Foundation of China (NSFC) under Grants No. 11474030 and No. U1530258, and by National Basic Research Program of China (973 Program) under Grant No. 2015CB921103, and by special program for applied research on super computation of the NSFC-Guangdong joint fund (the second phase).

\bibliographystyle{apsrev4-1}

\begin{thebibliography}{51}%

\bibitem{boykin2007} T. B. Boykin, N. Kharche, and G. Klimeck,
\newblock Phys. Rev. B {\bf 76}, 035310(2007).

\bibitem{boykin2007-2} T. B. Boykin, N. Kharche, G. Klimeck, and M. Korkusinski,
\newblock J. Phys.: Condens. Matter, {\bf 19}, 036203(2007).

\bibitem {popescu2010} V. Popescu, and A. Zunger,
\newblock Phys. Rev. Lett. {\bf 104}, 236403(2010).

\bibitem {ku2010} W. Ku, T. Berlijn, and C.-C. Lee,
\newblock Phys. Rev. Lett. {\bf 104}, 216401(2010).

\bibitem{ajoy2012} A. Ajoy, K. V. Murali, and S. Karmalkar,
\newblock J. Phys.: Condens. Matter {\bf 24}, 055504(2012).

\bibitem{lee2013} C.-C. Lee, Y. Yamada-Takamura, and T. Ozaki,
\newblock J. Phys.: Condens. Matter {\bf 25}, 345501(2013).

\bibitem {rubel2014} O. Rubel, A. Bokhanchuk, S. J. Ahmed, and E. Assmann,
\newblock Phys. Rev. B {\bf 90}, 115202(2014).

\bibitem{deretzis2014} I. Deretzis, G. Calogero, G. Angilella, and A. La~Magna,
\newblock  EPL (Europhys. Lett.) {\bf 107}, 27006(2014).

\bibitem {brommer2014} P. Brommer, and D. Quigley,
\newblock  J. Phys.: Condens. Matter {\bf 26}, 485501(2014).

\bibitem {zheng2015} F. Zheng, P. Zhang, and W. Duan,
\newblock  Comput. Phys. Commun. {\bf 189}, 213(2015).

\bibitem {farjam2015} M. Farjam,
\newblock  arXiv:1504.04937.

\bibitem {medeiros} P. V. C. Medeiros, S. Stafstr$\ddot{o}$m, and J. Bj$\ddot{o}$rk,
\newblock  Phys. Rev. B {\bf 89}, 041407(2014).


\bibitem{popescu2012} V. Popescu, and A. Zunger
\newblock Phys. Rev. B {\bf 85}, 085201(2012).

\bibitem {huang2014} H. Huang, F. Zheng, P. Zhang, J. Wu, B.-L. Gu, and W. Duan,
\newblock  New J. Phys. {\bf 16}, 033034(2014).

\bibitem {allen2013} P. B. Allen, T. Berlijn, D. Casavant, and J. Soler,
\newblock  Phys. Rev. B {\bf 87}, 085322(2013).


\bibitem {boykin2014} T. B. Boykin, A. Ajoy, H. Ilatikhameneh, M. Povolotskyi, and G. Klimeck,
\newblock  Phys. Rev. B {\bf 90}, 205214(2014).

\bibitem{huang2015} H. Huang,
\newblock J. Phys.: Condens. Matter {\bf 27}, 305402(2015).


\bibitem {qe} P. Giannozzi, S. Baroni, N. Bonini, M. Calandra, R. Car, C. Cavazzoni, D. Ceresoli, G. L. Chiarotti, M. Cococcioni {\it et. al.},
\newblock  J. Phys.: Condens. Matter {\bf 21}, {395502}(2009).


\bibitem {nc} D. Hamann, M. Schl{\"u}ter, and C. Chiang,
\newblock  Phys. Rev. Lett. {\bf 43}, 1494(1979).

\bibitem {pbe1} J. P. Perdew, K. Burke, and M. Ernzerhof,
\newblock  Phys. Rev. Lett. {\bf 77}, 3865(1996).





\end{thebibliography}

\end{document}